\font\twelvebf=cmbx12
\font\ninerm=cmr9
\nopagenumbers
\magnification =\magstep 1
\overfullrule=0pt
\baselineskip=18pt
\line{\hfil CCNY-HEP 96/2}
\line{\hfil RU-96-3-B}
\line{\hfil February 1996}
\vskip .3in
\centerline{\twelvebf On the origin of the mass gap for non-Abelian gauge}
\centerline{\twelvebf theories in (2+1) dimensions}
\vskip .4in
\baselineskip=14pt
\centerline{\ninerm DIMITRA KARABALI}
\vskip .05in
\centerline{Physics Department}
\centerline{Rockefeller University}
\centerline{New York, New York 10021}
\vskip .4in
\centerline{\ninerm V.P. NAIR}
\vskip .05in
\centerline{Physics Department}
\centerline{City College of the City University of New York}
\centerline{New York, New York 10031.}
\footnote{}{E-mail addresses: karabali@theory.rockefeller.edu, 
vpn@ajanta.sci.ccny.cuny.edu }
\vskip .5in
\baselineskip=16pt
\centerline{\bf Abstract}
\vskip .1in
An analysis of how the mass gap could arise in pure Yang-Mills theories in two spatial dimensions is given.
\vfill\eject
\footline={\hss\tenrm\folio\hss}
\magnification =\magstep 1
\overfullrule=0pt
\baselineskip=18pt
\pageno=2

\def\bp {\bar p}
\def\dag {\dagger}

\def\bdel{\bar{\partial}}

\def\bz {\bar{z}}
\def\12 {{\textstyle {1 \over 2}}}
\def\bG {\bar{G}}
\def\H {{\cal H}}
\def\vf {\varphi}
\def\S {{\cal S}}
\def\ra {\rangle}
\def\la {\langle}
\def\Tr {{\rm Tr}}

\def\bV{{\bar V}}
\def\bD {{\bar D}}
\def\bA {{\bar A}}
The study of non-Abelian gauge theories in two spatial dimensions, 
in particular the question of how a mass gap could arise in these theories, is 
interesting for at least two reasons: it is a useful guide to the more realistic case of 
three dimensions and secondly these theories can be an approximation to the high 
temperature phase of Chromodynamics with the mass gap serving as the magnetic mass.
In a recent paper, we considered the Hamiltonian analysis of (2+1)-dimensional gauge 
theories in a gauge-invariant matrix parametrization of the fields [1]. The kinetic term 
of the Hamiltonian, which is the Laplacian on the space of gauge-invariant configurations,
could be explicitly constructed in our parametrization. By considering eigenstates of the
Laplacian, one could then see
how a mass gap could arise in these theories. In this letter, 
we rederive the key results directly in terms of the gauge 
potentials and electric fields. A general expression for the Hamiltonian is also given
in terms of a current and derivatives with respect to it.

We consider the Hamiltonian version of an $SU(N)$-gauge theory in the $A_0=0$ gauge.
The gauge potential is written 
$A_i = -i t^a A_i ^a$, $i=1,2$, where $t^a$ are hermitian 
$N \times N$-matrices which form a basis of the Lie algebra of $SU(N)$ with
$[t^a, t^b ] = i f^{abc} t^c,~~{\Tr} (t^at^b) = {1 \over 2} \delta ^{ab}$. 
The Hamiltonian
can be written as
$$\eqalign{
\H& = T~+V \cr
T&={e^2\over 2} \int d^2 x ~ E_i^a E_i^a \cr
V&= {1\over 2e^2} \int d^2x~ B^a B^a \cr}
\eqno (1)
$$
where $e$ is the coupling constant and 
$B^a={1 \over 2} \epsilon_{jk}(\partial_j A_k^a - \partial_k A_j^a +f^{abc}
 A_j^b A_k^c )$.
We use complex coordinates $z=x_1 - i x_2, ~~ \bar{z} = x_1 +i x_2$ with the corresponding
components of the potential, viz., $A_{z} = {1 \over 2} (A_1 +i A_2), ~~ 
A_{\bar{z}} =
{1 \over 2} (A_1 -i A_2) = - (A_z)^{\dagger}$. The wavefunctions for the physical states
are gauge-invariant and have the inner product
$$
\la 1\vert 2 \ra ~=~ \int ~d\mu ({\cal C})~\Psi_1^* \Psi_2 \eqno(2)
$$
Here $d\mu ({\cal C})$ is the volume measure on the configuration space ${\cal C}$ which
is the space of gauge potentials ${\cal A}$ modulo the set of gauge transformations
${\cal G}_*$ which go to the identity at spatial infinity. The distance function on ${\cal A}$
is the standard Euclidean one
$$
ds^2~ =~ \int d^2x~ \delta A^a_i \delta A^a_i ~= -8 \int \Tr (\delta A_z \delta A_{\bz} )
\eqno(3)
$$
$d\mu ({\cal A})$ is the standard volume $[dA_z dA_{\bz} ]$ associated with (3) and
$d\mu ({\cal C})$ should be obtained by dividing out the volume of ${\cal G}_*$, i.e.,
$d\mu ({\cal C}) = [dA_z dA_{\bz} ]/vol({\cal G}_* )$.

The gauge potentials $A_z,~A_{\bz}$ can be parametrized in terms 
of complex $=SL(N,{\bf C})$-matrices $M,~M^\dag$ as
$$
A_z = -\partial_{z} M M^{-1},~~~~~~ A_{\bar{z}} = M^{\dagger -1} \partial_
{\bar{z}} M^{\dagger}
\eqno(4) 
$$
Given any gauge potential we can construct $M,~M^\dag$ at least as power series in the 
potential; for example,
$$\eqalign{
M(x) &= 1 - \int G(x,z_1 ) A(z_1) +\int G(x,z_1)A(z_1) G(z_1,z_2) A(z_2) - ...\cr
M^\dag (x) &=1 - \int \bA (z_1) \bG(z_1,x) +\int \bA (z_1) \bG(z_1,z_2) \bA (z_2) 
\bG(z_2,x) -...\cr
}\eqno(5)
$$
where $A\equiv A_z,~\bA \equiv A_{\bz}$ and $G,\bG $ are Green's functions for $\partial_z,~
\partial_{\bz}$ defined by
$$ 
\eqalignno{
{ \bar \partial}_x \bar{G} (x,y) ~= \partial _x &G(x,y) ~= \delta ^{(2)} (x-y) \cr
 \bar{G} (x,x') = {1 \over {\pi (z-z')}}, ~~~&~~~~~~
G(x,x') = {1 \over {\pi (\bar{z} - \bar{z}')}} &(6) \cr} 
$$ 
Notice that the solutions (5) may be summed up and written as
$$
\eqalign{
M (x)~&=~ 1 -\int_y  D^{-1}(x,y) A (y)\cr
M^\dag (x)~&=~1 - \int_y \bA  (y) \bD ^{-1} (y,x)\cr}\eqno(7)
$$
where $D=\partial +A,~\bD =\bdel +\bA$ are the covariant derivatives.

The matrix $M$ (or $M^\dag$) is not uniquely defined.
$M$ and $M {\bar V}(\bz )$,
where ${\bar V}(\bz )$ is antiholomorphic, and likewise $M^\dag$ and $V(z)M^\dag$, 
lead to the same potential. Eventually we must ensure that this ambiguity of parametrization 
does not affect physical results.

Under a gauge transformation, $A\rightarrow A^g =gAg^{-1} -dg~g^{-1}$ or equivalently,
$M\rightarrow gM,~M^\dag \rightarrow M^\dag g^{-1},~ g\in SU(N)$. $H=M^\dag M$ is 
gauge-invariant and this will be the basic field variable of the theory.

We first consider the evaluation of $d\mu ({\cal C})$. In terms of $M,M^\dag$, the metric
(3) becomes
$$
ds^2_{\cal A} ~= 8 \int \Tr \left[ D(\delta M M^{-1}) \bD (M^{\dag -1}\delta M^\dag )
\right] \eqno(8)
$$
(Here $D,\bD$ are in the adjoint representation.) The metric for $SL(N,{\bf C})$-matrices
is given by
$$
ds^2_{SL(N,\bf C)}  = 8 \int \Tr [(\delta M M^{-1}) (M^{\dagger -1} \delta M^{\dagger})] 
\eqno(9)
$$
The Haar measure $d\mu (M, M^{\dagger} )$ is the volume associated 
with this 
metric. The matrix $H$ belongs to $SL(N,{\bf C})/SU(N)$. The metric on this space becomes
$$
ds^2_H ~= 2 \int ~\Tr(H^{-1}~\delta H )^2 \eqno(10)
$$
From (8-10) we see that
$$
d\mu ({\cal C}) = {d\mu ({\cal{A}})\over vol({\cal G}_*)}
 = {[dA_z dA_{\bar{z}}]\over vol({\cal G}_*)} = (\det D_z D_{\bar{z}}) {d\mu 
(M, M^{\dagger})\over vol({\cal G}_*)}\eqno(11) 
$$
$d\mu(M,M^\dag )/vol({\cal G}_*)$ is given by the volume of the metric (10), viz.
$d\mu (H) = \det r [\delta \vf ]$ where $H^{-1}\delta H =\delta \vf^a r_{ab}t_b$,
$\vf^a$ being real parameters for the hermitian matrices $H$.
(A simple way to see this is the following.
$\rho \equiv (M^{\dag -1} d M^\dagger ~+ dM M^{-1})$ is a 
differential form on $SL(N,{\bf C})$ 
which transforms as
$\rho \rightarrow g \rho g^{-1}$ under $M\rightarrow g M$. 
Thus $\Tr ( \rho ^n)~= \Tr (H^{-1}dH )^n $ are differential forms on
$SL(N,{\bf C})/SU(N)$. The volume element is given by the differential form of maximal
degree, i.e., for $n= (N^2 -1)$. This is easily seen to be $\det r [d\vf ]$. 
For matrices which are functions of the spatial coordinates, as in our case, we have
the product over the spatial points as well, giving the result stated.)
With this result, we see from (11)
$$
d\mu ({\cal C}) ~=~ d\mu (H)~ \det (D \bD ) \eqno(12)
$$
The problem is thus reduced to calculating $\det (D \bD )$. The answer 
to this is well known, $\det (D \bD )= e^{2c_A \S (H)}$, $c_A \delta_{ab} =f_{amn}f_{bmn}$,
upto an irrelevant constant [2].
${\S} (H)$ is the Wess-Zumino-Witten (WZW) action for $H$.
$$
{\S} (H) = {1 \over {2 \pi}} \int \Tr (\partial H \bar{\partial} H^{-1})
+{i \over {12 \pi}} \int \epsilon ^{\mu \nu \alpha} \Tr ( H^{-1} \partial _{\mu}
H H^{-1} \partial _{\nu}H H^{-1} \partial _{\alpha}H) \eqno(13) 
$$
We thus have 
$$
d\mu ({\cal C}) = ~ d\mu (H)~ e^{2c_A \S (H) }  \eqno(14)
$$
This evaluation of $d\mu ({\cal C})$ has been given in reference [3].

Some of the details of the
calculation of $\det (D \bD )\equiv e^{\Gamma}$ are of interest in what
follows. From the definition
$$
{{\delta \Gamma}\over {\delta \bA^a(x)}}~=~(-i)\Tr \left[ \bD ^{-1}(x,y) 
T^a\right]_{y\rightarrow x}\eqno(15)
$$
where $(T^a)_{mn} = -i f_{amn}$. $\bD^{-1}(x,y)= M^{\dag -1}(x)M^\dag (y) \bG (x,y)$,
so that $\Tr [ \bD ^{-1}(x,y) T^a]_{y\rightarrow x}$ $ = -(c_A/\pi )(M^{\dag -1}\partial
M^\dag )^a$. This leads to $\Gamma = 2c_A {\cal S}(M^\dag )+f(M)$. This regularization 
is not gauge-invariant. For $d\mu ({\cal C})$, we need a gauge-invariant regularization of
 $\det (D \bD )$ such as covariant point-splitting or Pauli-Villars regulators.
$$
\eqalignno{
\bD^{-1} (x,y)_{Reg} &= \left[ \bD^{-1}(x,y) \exp (A(x-y)+\bA 
({\bar x}-{\bar y}))\right]_{y\rightarrow x} &(16a)\cr
&= \left[ \left( \bD^{-1} +D (\mu^2 -D \bD )^{-1}\right)(x,y)
\right]_{\mu^2 \rightarrow \infty} &(16b)\cr}
$$
Either one of these regulators gives
$$
\Tr \left[ \bD ^{-1} (x,y) T^a \right]_{y\rightarrow x}~= {1\over \pi} \Tr \left[
(A-M^{\dag -1}\partial M^\dag )T^a\right]\eqno(17)
$$
Correspondingly, we have ${\delta \Gamma \over \delta \bA^a}= 
(-ic_A/\pi )2 \Tr [t_a(A- M^{\dag -1}\partial M^\dag )]$, leading to $\Gamma =2c_A {\cal S}(H)$ and the result (14).  
We shall take (16a), viz. covariant point-splitting, as the regularization in what follows.

The inner product is now given by
$$
\la 1|2\ra 
= \int ~d\mu (H)~ e^{2c_A {\S} (H) }~~\Psi^*_1 \Psi_2 \eqno(18)
$$

In an intuitive sense, at this stage, we can see how a mass gap could possibly arise.
Writing $\Delta E, ~\Delta B$ for the root mean square fluctuations of the electric field $E$
and the magnetic field
$B$, we have, from the canonical commutation rules, $\Delta E~\Delta B\sim k$, 
where $k$ is the
momentum variable. This gives an estimate for the energy
$$
{\cal E}={1\over 2} \left( {e^2 k^2\over\Delta B^2 } +{\Delta B^2 \over e^2} \right) \eqno(19)
$$
For low lying states, we minimize ${\cal E}$ with respect to $\Delta B^2$, 
$\Delta B^2_{min}\sim
e^2 k$, giving ${\cal E}\sim k$. This is, of course, the standard photon or perturbative
gluon. However, for the non-Abelian theory, in calculating expectation values,
we must take account of
the factor $e^{2c_A \S }\approx
\exp [ -(c_A /2\pi ) \int B (1/k^2 )B +...]$. For low $k$,
this factor controls fluctuations in $B$, giving $\Delta B^2 \sim k^2 (\pi /c_A )$. 
In other words, eventhough ${\cal E}$ is minimized around $\Delta B^2 \sim k$, probability is 
concentrated around $\Delta B^2 \sim k^2 (\pi /c_A )$. For the expectation value of the energy,
we then find
${\cal E}\sim (e^2c_A/2\pi ) +{\cal O}(k^2)$. Thus the kinetic term in combination with 
the measure factor $e^{2c_A \S}$ could lead to a mass gap. (This argument is very similar to
how a mass is obtained for longitudinal plasma oscillations.) 
The argument is not rigorous; many
terms, such as the non-Abelian contributions to the commutators and $\S (H)$,
have been neglected.
Nevertheless, we expect this to capture the essence of how a mass gap could arise.

Matrix elements calculated with the inner product (18) are correlation functions of a
WZW-model for hermitian matrices. They can be calculated by analytic continuation
of the results for unitary matrices. For hermitian matrices, $e^{(k+2c_A)\S}$ is what 
corresponds to the action $e^{k \S (U)}$ for unitary matrices.
The 
analogue of the renormalized level $\kappa =k+c_A$ of the unitary WZW model is
$-(k+c_A)=-\kappa$. The correlators are obtained by the continuation
$\kappa \rightarrow -\kappa$. In the present case, we have only $e^{2c_A\S}$, 
hence we must also take $k\rightarrow 0$. 
In the unitary case, correlators involving primary fields belonging to the 
nonintegrable representations of the current algebra vanish. The corresponding statement
for the hermitian model is that such correlators become infinite or undefined.
For $k\rightarrow 0$, only the identity and its current algebra
descendants have 
well-defined correlators. (The divergence of correlators of other operators
has to do with $k\rightarrow 0$, not coincidence of arguments.)
We must thus conclude that the wavefunctions can be taken as
functions of the current 
$(\partial H ~H^{-1})_a $.

The above arguments give the conformal field-theoretic reason for the currents
being the quantities of interest. The result, however, is not surprising since the
Wilson loop operator can be written in terms of $H$ as
$$
W(C)~=~ \Tr~ P \exp\left( - \oint_C dz~\partial H H^{-1}\right) \eqno(20)
$$
The Wilson loop operators form a complete set and hence the currents should suffice 
to generate the gauge-invariant states.

The vacuum state for the kinetic term is given by $\Psi_0 =$constant. This is normalizable 
with the inner product (18). From the above arguments, the quantity of interest in 
constructing higher states is the
current
$$\eqalign{
J_a(x) &= {c_A \over \pi } \left( \partial H~H^{-1} \right)_a (x) \cr
&= {c_A \over \pi} \left[ iM^{\dag }_{ab} (x) A_b(x) + (\partial M^\dag ~M^{\dag -1})_a (x)
\right] \cr
}\eqno(21)
$$
where $M^{\dag }_{ab}= 2 \Tr (t_a M^\dag t_b M^{\dag -1} )$ is the adjoint 
representation of $M^\dag$. The action of the kinetic energy operator can be calculated as
follows.
$$\eqalign{
T~\Psi (J) ~&= -{e^2\over 2} \int_z {\delta^2 \Psi \over \delta {\bar A}_k(z) \delta A_k(z)}\cr
&= -{e^2\over 2}\left[ \int_{x,y,z} {\delta J_a(x) \over \delta A_k(z)}{\delta J_b(y)
\over \delta {\bar A}_k(z)} {\delta^2 \Psi \over \delta J_a(x) \delta J_b(y) }
+ \int_{x,z} {\delta^2 J_a(x) \over \delta {\bar A}_k(z) \delta A_k(z) }{\delta \Psi \over
\delta J_a(x)}\right]\cr} \eqno(22)
$$
From the definition of $J_a$ and $M^\dag$ we get
$$\eqalignno{
{\delta J_a(x) \over \delta A_k(z)}&= i{c_A\over \pi}M^\dag_{ak}(x) \delta (x-z)&(23a)\cr
{\delta J_b(y) \over \delta {\bar A}_k(z)}&= -i{c_A\over \pi}\left( {\cal D}_y \bG (y,z)
M^\dag (z)\right)_{bk}&(23b)\cr
{\cal D}_{mn}&= \partial \delta_{mn} +i{\pi \over c_A} f_{mnc}J_c &(23c)\cr}
$$
Using (23a) and the definition of $M^\dag$, we find
$$\eqalign{
\int_z{\delta^2 J_a(x) \over \delta {\bar A}_k(z) \delta A_k(z) }&= i{c_A \over \pi}
\left[ {\delta M^{\dag}_{ab}(x) \over 
\delta \bA_b(y)}\right]_{y\rightarrow x} \cr
&={c_A\over \pi} M^\dag _{am} \Tr \left[ T^m \bD ^{-1}(y,x) 
\right]_{y \rightarrow x}\cr}\eqno(24)
$$
The coincident limit of $\bD^{-1}$ which appears in this equation has to be evaluated by
regularization.The arguments are the same because both functional derivatives act at the same
point. We can consider the kinetic energy operator as $E^k_z(z)E^k_{\bz}(z'),~z'\rightarrow
z$. This would give a point-splitting regularized version of $\bD^{-1}$. However, it is not
gauge-invariant; we need a phase factor connecting the two points $z$ and $z'$. The
covariantly regularized expression can be written as
$$\eqalign{
T~&= 2e^2 \int_z \left[ E^k_z (z) P(z,z')^{kl} E^l_{\bz}(z')\right]_{z'\rightarrow z}\cr
P(z,z')&= \exp\left[ -A (z-z') -\bA (\bz -\bz ')\right]\cr}\eqno(25)
$$ 
In $P(z,z')$, the potentials $A$ and $\bA$ are evaluated at $\12 (z+z')$; this 
midpoint specification is consistent with the hermiticity of $T$.
Notice also that, in this case, $E_z(z) P(z,z') = P(z,z') E_z(z)$. Using (25), we get
$$
\int_z \left[ {\delta^2 J_a(x) \over \delta {\bA}_k(z) \delta A_k(z)} \right]_{Reg}
= {c_A\over \pi} M^\dag_{am} \Tr \left[ T^m \bD^{-1} (z,x)P^T (z,x)\right]_{z\rightarrow x}
\eqno(26)
$$
($P^T$ is the transpose of $P$.)
We see that the use of expression (25) for $T$ is equivalent to covariant point-splitting
regularization of $\bD^{-1}$. Using (17), we finally get
$$
\int_z ~{\delta^2 J_a(x) \over \delta {\bA}_k(z) \delta A_k(z)} = -{c_A\over \pi} J_a(x) 
\eqno(27)
$$
Combining (22,23,27), the kinetic energy operator $T$ is obtained as
$$\eqalign{
T ~\Psi &= m \left[ \int_x J_a(x) {\delta \over \delta J_a(x)} +
\int_{x,y}\Omega_{ab}(x,y) {\delta \over \delta J_a (x)} 
{\delta \over \delta J_b(y)}\right]\Psi \cr 
\Omega_{ab}(x,y) &= \left[ {c_A\over \pi }\delta_{ab} \partial_y \bG (x,y) -i f_{abc} J_c(y)
\bG (x,y) \right] \cr}
\eqno(28)
$$
where $m= e^2c_A/2\pi $. In particular, we see that $J_a$ is an eigenfunction of
$T$ with eigenvalue $m$, i.e., 
$$
T~J_a(x) = {e^2 c_A \over 2\pi }~J_a(x) \eqno(29)
$$
Of course,
$J_a$ by itself would not be an acceptable eigenfunction since it is not invariant under
$H \rightarrow V(z) H{\bar V}({\bar z})$. We have to construct suitable combinations
of $J_a$'s. Nevertheless, the result (29) is the mathematical expression of the intuitive
arguments given earlier. (Expression (28) is also typically of the form which arises in change of
variables or the introduction of collective coordinates in field theory [4].) 

The ambiguity in defining $M$ for a given potential $A$ is a constant matrix $\bV$ if we
use boundary conditions on $M$ appropriate to a Riemann sphere. However, from the point of
view of constructing $M$ from $A$, there is no reason why this should not be done 
independently in different regions of space with matching conditions on overlap 
regions. Thus we must allow the freedom of making transformations $H \rightarrow
V(z) H \bV (\bz )$ for the purpose of matching $H$'s in different regions. 
Proper wave functions are thus constructed from products of the currents 
by requiring this invariance as well. (This point was not elaborated upon in [1].)

The wavefunction for the simplest excited state is given by the product of 
two currents. We find, using (28), that
$$
\alpha = J_a(x) J_a (y) + {c_A{\rm dim}G \over \pi^2} {1\over (x-y)^2} \eqno(30)
$$
is orthogonal to the ground state and is an eigenfunction of $T$ with eigenvalue $2m$.
(dim$G$ is the dimension of the group $G=SU(N)$.)
By applying $\bdel _x,~\bdel_y$ and taking $y \rightarrow x$, we can construct a state
$\Psi_2$
of eigenvalue $2m$, which
is invariant under
$H \rightarrow V(z) H \bV (\bz )$. 
$$
\Psi_2 (J) = \int_x f(x) \left[ \bdel J_a(x) \bdel J_a(x) + {c_A{\rm dim}G \over \pi^2} 
\partial_x
\bdel _x \delta (x-y) \vert_{y\rightarrow x} \right] \eqno(31)
$$
The second (c-number) term in (31) orthogonalizes 
this with respect to the ground state. (One may regard $\bdel J_a(x) \bdel J_a(y)$ as 
providing a point-split version of $B_a(x) B_a(x)$. A point-splitting respecting invariance
under $H \rightarrow V(z) H \bV (\bz )$ would be 
$$
\beta (x,y) = B_a(x)[P \exp( -\int_y^x \partial H H^{-1} )]_{ab} B_b(y) \eqno(32)
$$
In the limit $y \rightarrow x$, $\beta (x,y)$ gives (29) and $T\beta$ goes to $T \Psi_2$
as well.) The construction of higher excited states will be discussed elsewhere.

In ref.[1],
an expression for $T$ was given in terms of derivatives with respect to the parameters
$\vf^a$ of $H$ as defined after (11).
The matrix element of the kinetic energy term is given by
$$
\la 1|T|2\ra = {e^2\over 2} \int 
 d\mu (H) e^{2 c_A {\S} (H)}\int_x \left[(Gp_a \Psi_1)^* K_{ab} 
(G p_b \Psi_2)  \right]  
\eqno(33)
$$
where
$$\eqalign{
p_m = -i r^{-1}_{mn} {\delta \over \delta \vf^n},~~~~~~&~~~~~~{\bp}_m =-ir^{*-1}_{mn}
{\delta \over \delta \vf^n}\cr
K_{ab}= 2\Tr &(t_a H t_b H^{-1} )\cr}
\eqno(34)
$$
The action of $T$ on products of $J$'s can also be computed from this. Taking $\Psi_2$ to be
a function of the current, we get
$$\eqalign{
\la 1|T|2\ra &= i{e^2 c_A\over 2\pi} \int 
 d\mu (H) e^{2 c_A {\S} (H)}\int_x \left[(Gp_a \Psi_1)^* K_{ab} 
 K_{cb}{\delta \Psi_2 \over \delta J_c }\right] \cr
&= im \int d\mu (H)~ \left[ \Psi_1^* \int_x \bG \bp_a \left\{ e^{2c_A \S (H)} 
{\delta \Psi_2 \over \delta J_a } \right\} \right]\cr
&= im \int d\mu (H) e^{2c_A \S (H)}~ \left[ \Psi_1^* \int_x \left\{ -iJ_a(x) +\int_y 
[(\bG \bp_a)(x) J_b(y)] {\delta \over \delta J_b(y)}\right\}
{\delta \Psi_2 \over \delta J_a(x) }  \right]\cr}\eqno(35)
$$
We have used the relation $(Gp_b)(x)J_c(y) = (ic_A/\pi )K_{cb}\delta (x-y)$ and
$(\bG \bp_a)\S =(-i/2\pi)(\partial H H^{-1})_a$. Using also $(\bG \bp_a)(x) J_b(y)
= -i \Omega_{ab}(x,y)$, we see that (35) leads to exactly the same expression as (28). 
Since it suffices, by our earlier arguments, to consider only wave functions
which are functions of the currents, it follows that the operator $T$ as given by (33) 
is the same as (25) or (28),
giving an alternative confirmation of the calculations in [1].
Further, $T$ as given by (33) is evidently self-adjoint;
thus (28) is self-adjoint as well, despite the naive lack of manifest hermiticity.
In collective field theory, rather than demonstrating self-adjointness, one usually
determines the measure factor appearing in the inner product
by requiring self-adjointness of the Hamiltonian [4]. It is clear from the above 
calculations that the measure so determined will lead to the inner product (18).

It is interesting to note that the
potential energy can also be written in terms of $J_a$'s as
$$
V ~\Psi = \left[ {1\over m} {\pi \over c_A} \int_x \bdel J_a (x) \bdel J_a(x)\right] 
~\Psi\eqno(36)
$$
\vskip .1in
We thank G. Alexanian, R. Jackiw, B. Sakita, S. Samuel and 
especially Chanju Kim for useful
discussions. This work was supported in part by the Department of Energy, grant
number DE-FG02-91ER40651-Task B and the National Science
Foundation, grant number PHY-9322591.
\vskip .2in
\noindent{\bf References}
\vskip .1in
\item{1.}
D. Karabali and V.P. Nair, Preprint hep-th / 9510157, October 1995 (to
be published in {\it Nucl.Phys. B}).
\item{2.}
A.M. Polyakov and P.B. Wiegmann, {\it Phys.Lett.} {\bf 141B} (1984) 223;
D.Gonzales and A.N.Redlich, {\it Ann.Phys.(N.Y.)} {\bf 169}, 104 (1986);
B.M. Zupnik, {\it Phys.Lett.} {\bf B183}, 175 (1987).
\item{3.}
K. Gawedzki and A. Kupiainen, {\it Phys.Lett.} {\bf 215B} (1988) 119;
{\it Nucl.Phys.} {\bf B320} (1989) 649.
\item{4.}
see, for example, B. Sakita, {\it Quantum theory of many variable systems and fields}
(World Scientific, 1985).

\end